\PassOptionsToPackage{unicode}{hyperref}
\PassOptionsToPackage{hyphens}{url}
\documentclass[11pt]{article}
\usepackage[a4paper,margin=1in]{geometry}

\usepackage{iftex}
\usepackage[T1]{fontenc}
\usepackage[utf8]{inputenc}
\usepackage{textcomp}
\usepackage{lmodern}
\usepackage{amsmath,amssymb}
\usepackage{lmodern}

\usepackage{xcolor}
\usepackage{amsmath,amssymb}
\usepackage{parskip}
\usepackage{calc}
\usepackage{array}
\usepackage{booktabs}
\usepackage{longtable}
\usepackage{multirow}
\usepackage{graphicx}
\usepackage{placeins}
\usepackage{caption}
\usepackage{hyperref}

\setkeys{Gin}{width=\linewidth,keepaspectratio}
\newcommand{\pandocbounded}[1]{#1}

\providecommand{\tightlist}{%
  \setlength{\itemsep}{0pt}\setlength{\parskip}{0pt}}

\hypersetup{
  hidelinks,
  pdftitle={EastAsiaClimateExtremes: An AI-Ready Dataset of Weekly Atmospheric and Oceanic Extremes over East Asia for Subseasonal Prediction Research},
  pdfauthor={Miae Kim, Yun-Young Lee, Uran Chung}}

\begin{document}

\title{EastAsiaClimateExtremes: An AI-Ready Dataset of Weekly
Atmospheric and Oceanic Extremes over East Asia for Subseasonal
Prediction Research}

\author{%
Miae Kim,
Yun-Young Lee,
Uran Chung%
}

\date{%
{\small Prediction Research and Development Department, Climate Services and Research Division, APEC Climate Center, Republic of Korea}\\[3pt]
\footnotesize Author email(s): \texttt{miaekim@apcc21.org},
\texttt{dolkong400@gmail.com},
\texttt{uchung@apcc21.org}%
}

\maketitle

\begin{abstract}

Despite growing interest in AI-based prediction of climate extremes,
event- or label-based AI-ready extreme climate datasets remain limited,
constraining efforts to systematically characterize and forecast such
phenomena. To address this gap, we present EastAsiaClimateExtremes, an
open dataset that provides ERA5/OISST reanalysis-based weekly extreme
labels and event-based metrics for anomalously high temperature (AHT),
heavy rainfall (HR), and marine heatwaves (MHW) over East Asia, together
with analysis workflows hosted on GitHub to facilitate reproducibility
and adaptation. The dataset is fully documented and co-registered with
ECMWF S2S hindcast outputs on a common spatial grid and temporal
framework, thereby enabling direct comparison between reanalysis-derived
labels and dynamical model forecasts. This unified dataset serves as a
reference framework for East Asian climate extreme research and AI-based
subseasonal-to-seasonal prediction. It supports both quantitative
characterization of the spatiotemporal occurrence of regional extremes
and systematic diagnosis of the S2S model skill in reproducing extreme
signals. Beyond these immediate applications, the dataset enables a
broader range of studies such as extreme event attribution and compound
risk analysis. The accompanying analysis workflows characterize the
historical statistics of reanalysis-based weekly extremes---including
occurrence frequency and mean and maximum intensity---together with
their climatological means and trend characteristics, and further assess
the skill of the ECMWF hindcast. Consistent with ERA5, both V2016 and
V2024 hindcast versions show that the frequency of extreme occurrence
has increased substantially during the analysis period. However, in particular,
the spatial distribution of the frequency of AHT occurrence over the 20-year
period is considerably displaced from the reanalysis, particularly in
V2024. This spatial discrepancy highlights the limitations of dynamical
models in representing the regional patterns of extreme events.

\end{abstract}

\section{1. Introduction}\label{introduction}

The East Asian region is one of the most densely populated and
economically active areas in the world, and has repeatedly suffered
substantial socio-economic damage from diverse extreme climate events.
Such events can lead to a range of impacts including increased
heat-related health risks, floods and landslides, and threats to marine
ecosystems and associated fisheries losses (Vicedo-Cabrera et al., 2021;
Smale et al., 2019). As their frequency and intensity are increasing
under ongoing climate change, strengthening the capability to predict
extreme events on subseasonal to seasonal timescales has emerged as a
key challenge (Vitart and Robertson, 2018; Domeisen et al., 2022).

To address this challenge, numerous studies have developed
subseasonal-to-seasonal prediction systems and artificial intelligence
(AI)-based forecasting models using observational, reanalysis, and
numerical model data. The Subseasonal-to-Seasonal (S2S) prediction
project was launched to support forecasts from two weeks to several
months ahead. The release of various climate model datasets, such as the
ECMWF S2S prediction system, has substantially expanded the basis for
research on extreme event prediction (Vitart et al., 2017). Previous
studies have evaluated ensemble-based temperature and precipitation
forecasts and found statistically significant subseasonal predictability
at subseasonal lead times (Pegion et al., 2019; Li and Robertson, 2015;
Wang and Robertson, 2019). Building on this predictive potential,
AI-based subseasonal-to-seasonal prediction models have actively been
developed. AI-based approaches have predominantly focused on predicting
continuous mean fields (e.g., precipitation, temperature) from
reanalysis or numerical model outputs; statistically post-processing and
bias-correcting existing forecasts; or performing regression on scalar
climate indices defined over large-scale domains (Pathak et al., 2022;
Bi et al., 2023; Lam et al., 2023; Kochkov et al., 2024; Price et al.,
2024; Lang et al., 2024; Bodnar et al., 2025; Lang et al., 2026).
Nevertheless, increasing interest in climate extreme prediction using
AI, AI-ready, event/label-based extreme climate datasets remain limited.
Furthermore, predicting the occurrence and intensity of extreme events
with sufficient reliability remains a major challenge.

Despite these developments, there remains a structural gap between
current approaches and extreme event labels that AI models can directly
learn from, particularly for spatially localized and temporally
intermittent extreme events on weekly timescales. To bridge this gap, it
is essential to have consistently labelled gridded extreme datasets that
can be used for model training and verification; however, public
resources specifically tailored to this purpose are scarce. For example,
existing extreme climate analysis products such as the Expert Team on
Climate Change Detection and Indices (ETCCDI)-type indices or the
Generalized Extreme Value (GEV) distribution and pointwise extreme value
(PEV) statistics are useful for diagnosing long-term changes, but
generally they are not directly converted into grid-based binary or
probabilistic labels or into event-level information required for weekly
prediction problems. Recently, several AI-oriented climate prediction
benchmarks have been proposed, including WeatherBench/WeatherBench-2,
which evaluate general raw and derived variables at the global scale
(Rasp et al., 2020; Rasp et al., 2024), and Extreme Weather Bench, which
compiles a global collection of individual high-impact case
studies---such as heatwaves, large-scale cold events, severe convective
outbreaks, atmospheric rivers, and tropical cyclones---to benchmark
forecasts against selected historical events (McGovern et al., 2026). In
contrast, EastAsiaClimateExtremes provides continuous, grid-point-level
weekly extreme labels and indices across the full historical record for
East Asia, co-registered with ECMWF S2S hindcasts, thereby supporting
systematic training, diagnosis, and skill assessment of extreme event
prediction models at a regional scale. Furthermore, as the criteria for
defining extreme climate events vary across event types, converting raw
data into a grid-level format suitable for AI models requires
substantial preprocessing effort and computational cost. To alleviate
this burden and enhance the usability of climate extreme data for
research, it is necessary to provide a dataset that can be directly used
as labels in a directly usable form, together with analysis tools that
allow users to easily explore the statistical characteristics of extreme
climate events.

In this study, we present EastAsiaClimateExtremes, an open dataset for
characterizing and analyzing climate extremes over East Asia, publicly
available on GitHub. This dataset provides weekly extreme labels and
event-based metrics for anomalously high temperature (AHT), heavy
rainfall (HR), and marine heatwaves (MHW) on a \(1.5^\circ \times 1.5^\circ\) grid over East Asia. Extreme events are defined using seasonally adjusted
percentile-based thresholds and minimum persistence criteria, enabling
the analysis of representative cases and statistical properties of
extreme climate events over East Asia. In addition, the repository
includes example code for generating weekly and monthly extremeness
maps, time series and heatmap visualizations, and long-term trend
analyses based on reanalysis data, as well as for extracting extreme
events from ECMWF hindcasts and comparing their statistics against
reanalysis-derived extreme labels. These resources provide a practical
starting point for systematically evaluating and improving subseasonal
extreme event prediction performance with diverse AI or statistical
models. In the following subsections, we summarize the main components
of the repository, focusing on (A) the overview of the extreme climate
datasets and (B) example code for extreme event statistics and
diagnostics.

\section{2. Repository Overview and
Workflow}\label{repository-overview-and-workflow}

The EastAsiaClimateExtremes repository hosts the open dataset and
accompanying analysis workflows described in this paper, publicly hosted
on GitHub. It is organized to make it easy for users to access the
datasets, reproduce the main results, and adapt the provided scripts to
their own applications. The following subsections describe the main
components of the repository, focusing on (A) climate extreme datasets
and (B) example codes for extreme event statistics and diagnostics.

\subsection{A. Overview of Extreme Climate Datasets}

Figure 1 shows a conceptual workflow of the EastAsiaClimateExtremes
dataset. In the upper workflow, reanalysis data are regridded and
temporally aggregated. Extreme event definitions are then applied to
derive historical weekly extreme labels and event-based metrics. The
reanalysis-based products provide a reference for evaluating extreme
event statistics from dynamical models and can also serve as inputs or
targets for AI-based prediction. A parallel workflow is applied to ECMWF
hindcasts as shown in the lower part of Figure 1. Model outputs are
preprocessed and spatiotemporally aligned with the observational
reference data, and then transformed into weekly extreme labels and
event-based metrics. The resulting hindcast products can be used as
dynamical benchmark forecasts for extreme events and enable a consistent
evaluation of AI-based extreme prediction models.

\begin{figure}
\centering
\pandocbounded{\includegraphics[keepaspectratio,alt={Figure 1. Conceptual workflow of constructing the EastAsiaClimateExtremes datasets.}]{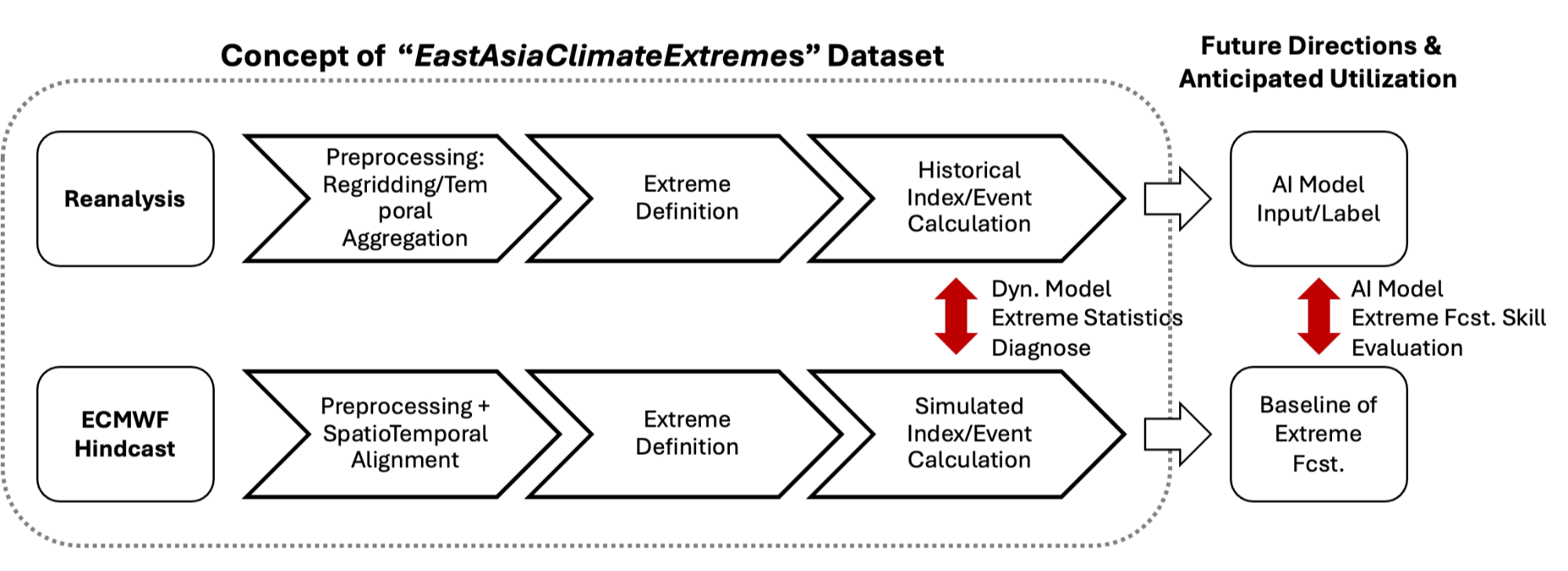}}
\caption{Conceptual workflow of constructing the
EastAsiaClimateExtremes datasets.}
\label{fig:workflow}
\end{figure}

The repository data directory is organized as follows:

\begin{verbatim}
DataFilesforEastAsiaClimateExtremes
|-- 0.ExtremeEvents_ERA5_OISST
|-- 1.Daily_ERA5_OISST
|-- 2.Weekly_ERA5
\-- 3.Weekly_ECMWFhcst
\end{verbatim}

The data folders 1, 2, and 3 contain daily (1.Daily\_ERA5\_OISST) and
weekly (2.Weekly\_ERA5 and 3.Weekly\_ECMWFhcst) climate variables along
with their quantile-based extreme thresholds. Table 1 summarizes the
daily and weekly climate data available in the folders. The dataset
includes three key variables---2-m air temperature (T2M), sea surface
temperature (SST), and total precipitation (TP)---derived from ERA5,
OISST (SST only), and ECMWF hindcast products. For each variable, three
types of data are provided: the original time series, the climatological
mean, and the 90th and 95th percentile thresholds (P90 and P95) on a
daily and weekly basis. Folder 0 contains weekly (7day rolling) and
monthly extremeness metrics including extreme days, maximum intensity,
and impact factor for the three variables, as detailed in Table 2. Further, 
it also contains extreme event properties for each variable and
data source, including the total number of identified events over the
analysis period, event start and end dates, duration, and other event
characteristics with extreme event duration--gap criteria used for event
identification, as shown in Table 3.

{\def\LTcaptype{table}
\begin{longtable}[]{@{}
  >{\raggedright\arraybackslash}p{(\linewidth - 2\tabcolsep) * \real{0.2000}}
  >{\raggedright\arraybackslash}p{(\linewidth - 2\tabcolsep) * \real{0.8000}}@{}}

\caption{Description of daily and weekly climate data available in the
\texttt{EastAsiaClimateExtremes} repository.}
\label{tab:daily_weekly_climate_data}\\

\toprule\noalign{}
\textbf{} & \textbf{Description} \\
\midrule\noalign{}
\endfirsthead

\toprule\noalign{}
\textbf{} & \textbf{Description} \\
\midrule\noalign{}
\endhead

\bottomrule\noalign{}
\endlastfoot
Type & original time series, climatological long-term mean, and 90th/95th
percentile thresholds (P90/P95) \\
Variables & 2\,m air temperature (T2M), sea surface temperature (SST),
and total precipitation (TP) \\
Resources & ERA5, OISST (SST only), and ECMWF hindcasts \\
Location &
\texttt{EastAsiaClimateExtremes/DATA/1.Daily\_ERA5\_OISST/},
\texttt{2.Weekly\_ERA5/}, and
\texttt{3.Weekly\_ECMWFhcst/} \\
\end{longtable}
}

{\def\LTcaptype{table}
\begin{longtable}[]{@{}
  >{\raggedright\arraybackslash}p{(\linewidth - 2\tabcolsep) * \real{0.28}}
  >{\raggedright\arraybackslash}p{(\linewidth - 2\tabcolsep) * \real{0.72}}@{}}

\caption{Definition of weekly and monthly extremeness metrics.}
\label{tab:extremeness_metrics}\\

\toprule\noalign{}
\textbf{Metric} & \textbf{Definition} \\
\midrule\noalign{}
\endfirsthead

\toprule\noalign{}
\textbf{Metric} & \textbf{Definition} \\
\midrule\noalign{}
\endhead

\bottomrule\noalign{}
\endlastfoot

\textbf{Extreme Days (ED)} &
Number of days within a 7-day window or calendar month for which the
daily variable exceeds its seasonally adjusted extreme threshold. \\

\textbf{Maximum Intensity (MI)} &
Maximum daily exceedance above the seasonally adjusted extreme threshold
within a 7-day window or calendar month. \\

\textbf{Impact Factor (IF)} &
Cumulative daily departure from the seasonally adjusted extreme threshold
within a 7-day window or calendar month. \\

\end{longtable}
}

{\def\LTcaptype{table}
\begin{longtable}[]{@{}
  >{\raggedright\arraybackslash}p{(\linewidth - 2\tabcolsep) * \real{0.2000}}
  >{\raggedright\arraybackslash}p{(\linewidth - 2\tabcolsep) * \real{0.8000}}@{}}

\caption{Definition of event properties and event-detection criteria.}
\label{tab:event_properties_criteria}\\

\toprule\noalign{}
\textbf{} & \textbf{Description} \\
\midrule\noalign{}
\endfirsthead

\toprule\noalign{}
\textbf{} & \textbf{Description} \\
\midrule\noalign{}
\endhead

\bottomrule\noalign{}
\endlastfoot

\textbf{Event properties} &
\textbf{N\_events}: Total number of detected events for individual events for study area, as applicable [A scalar integer]. \newline
\textbf{Start date / end date}: Start and end dates of an event
[datetime]. \newline
\textbf{Duration}: Length of an event [days]. \newline
\textbf{Mean intensity}: Mean intensity [deg. C].
\newline
\textbf{Peak intensity}: Maximum (peak) threshold [deg. C].
\newline
\textbf and so forth \\

\textbf{Event-detection criteria} &
\textbf{D3G5*} and \textbf{D5G2} for AHT and MHW,
\textbf{D1G3} and \textbf{D3G3} for HR. *denotes minimum 3-day duration, maximum 5-day gap. \\

\end{longtable}
}

\subsection{B. Analysis Codes for Extreme Event Statistics and
Diagnostics}\label{b.-analysis-codes-for-extreme-event-statistics-and-diagnostics}

The repository also includes a set of example analysis workflows for
extreme event statistics and diagnostics, designed to help users explore
both time series and spatial patterns of AHT, HR, and MHW (Table 4). The
first notebook, \path{Extreme_Event_Statistics_and_Visualization_1.ipynb},
visualizes extreme events at a user-selected grid point over East Asia.
Users can specify the event type, percentile threshold, duration--gap
criterion, target coordinates, and analysis periods. The notebook loads
the precomputed event records together with daily data, climatological
means, and threshold data. It identifies the nearest grid point to the
selected location and reconstructs a binary daily event series from the
recorded event start and end dates.

Three visualizations are produced:

\begin{enumerate}
\def\labelenumi{\arabic{enumi}.}
\tightlist
\item
  Daily time series: Displays the original daily variable,
  climatological mean, and percentile threshold. Periods classified as
  extreme events are highlighted.
\item
  Annual event-day counts: Aggregates the binary daily series by year
  and estimates a linear trend.
\item
  Monthly heatmap: Shows monthly event-day counts for each year. The
  sign of the linear trend is indicated for each calendar month.
\end{enumerate}

The notebook,
\path{Extreme_Event_Statistics_and_Visualization_2.ipynb}, calculates and
visualizes long-term statistics of extreme events over East Asia. Users
can select the target event type, percentile threshold, duration--gap
criterion, data source (ERA5 or OISST), and analysis period. For each
year and grid point, the script calculates event records from the
precomputed event dataset and three statistics: 1) mean annual event
frequency, 2) mean annual total event duration, and 3) mean annual event
intensity. It then estimates long-term linear trends in these annual
metrics using ordinary least squares regression. Trend values are
expressed per decade. 
The notebook,
\path{Weekly_Extreme_Statistics_and_Visualization.ipynb}, calculates and
visualizes climatological weekly extreme statistics for AHT, HR, and MHW
over East Asia. The analysis covers 1940--2024, while extreme thresholds
are defined using a 1991--2020 climatological reference period. Users
can select the target variable, percentile threshold, climatological
reference period, and spatial domain. The notebook loads precomputed
7-day rolling mean or accumulated fields and their day-of-year-dependent
percentile thresholds. For visualization, we exclusively extract weekly
samples initialized on Mondays (at 7-day intervals), defining an extreme
week as one where this 7-day value exceeds its corresponding threshold.

For each grid point, three statistics are calculated:

\begin{itemize}
\tightlist
\item
  Mean annual frequency: the average number of extreme weeks per year.
\item
  Mean annual extreme-week intensity: for each year, the mean intensity
  across extreme weeks; the climatological value is the average of this
  annual metric over the analysis period.
\item
  Mean annual maximum intensity: the annual maximum among extreme weeks,
  averaged across the analysis period.
\end{itemize}

Finally, the notebook,
\path{Weekly_Extreme_Statistics_and_Visualization_ECMWF.ipynb}, compares
weekly extreme statistics from ECMWF S2S hindcasts with ERA5 reanalysis
data over East Asia for AHT, HR, and MHW. The ECMWF hindcast datasets
are extracted from the 2016 and 2024 ECMWF hindcast versions. For each
initialization date, the notebook extracts lead week 3 (forecast days
15--21) data and compares them with ERA5 7-day rolling fields over the
corresponding valid period. Hindcast samples are selected according to
the ECMWF initialization schedule available for each version, and only
initialization dates with matching ERA5 valid-period data are retained.
Extreme conditions are defined as weekly values exceeding a
day-of-year-dependent 90th-percentile threshold. Version-specific
thresholds are calculated separately using the corresponding hindcast
initialization dates for each ECMWF version. These thresholds allow each
hindcast version to be assessed relative to its own climatology, while
the corresponding ERA5 subsets help distinguish period-dependent changes
from model--reanalysis differences. The notebook calculates the mean
annual frequency of threshold exceedances for:

\begin{itemize}
\tightlist
\item
  ECMWF hindcast version 2016 over 1996--2015, with a version-specific
  week-3 threshold.
\item
  ECMWF hindcast version 2024 over 2004--2023, with a version-specific
  week-3 threshold.
\item
  For comparison, ERA5 subsets for 1996--2015 and 2004--2023, using
  corresponding threshold climatologies.
\end{itemize}

In addition, the notebook generates a grid-based timeseries comparison
for a user-selected location. For a representative year
(e.g., 2023), it plots ECMWF version-2024 week-3 values and ERA5 values
together with their respective thresholds. The areas where the timeseries exceed the thresholds are shaded separately for the hindcast and reanalysis data.

{\def\LTcaptype{table}
\begin{longtable}[]{@{}
  >{\raggedright\arraybackslash}p{(\linewidth - 2\tabcolsep) * \real{0.3400}}
  >{\raggedright\arraybackslash}p{(\linewidth - 2\tabcolsep) * \real{0.6600}}@{}}

\caption{Summary of analysis workflows included in the
\texttt{EastAsiaClimateExtremes} repository.}
\label{tab:analysis_workflows}\\

\toprule\noalign{}
\textbf{Code file} & \textbf{Description} \\
\midrule\noalign{}
\endfirsthead

\toprule\noalign{}
\textbf{Code file} & \textbf{Description} \\
\midrule\noalign{}
\endhead

\bottomrule\noalign{}
\endlastfoot

\path{Extreme_Event_Statistics_and_Visualization_1.ipynb} &
Visualizes daily extreme-event time series, annual event-day counts,
linear trends, and monthly event-day heatmaps at a user-selected grid
point. \\

\path{Extreme_Event_Statistics_and_Visualization_2.ipynb} &
Calculates and visualizes spatial maps of long-term extreme-event
frequency, duration, intensity, and their trends over East Asia. \\

\path{Weekly_Extreme_Statistics_and_Visualization.ipynb} &
Calculates and visualizes maps of climatological weekly extreme
frequency, mean extreme-week intensity, and mean annual maximum
intensity. \\

\path{Weekly_Extreme_Statistics_and_Visualization_ECMWF.ipynb} &
Compares lead-week-3 extreme statistics from ECMWF S2S hindcasts with
corresponding ERA5 reference products, including spatial maps and
point-based time-series diagnostics. \\

\end{longtable}
}

Section 3 summarizes the original datasets utilized in this study. The
resultant data products, introduced in Section 2, are organized by
variable and data source and are further detailed in Section 4. These
products are accompanied by analysis workflows, facilitating
reproducible statistical analyses of East Asian climate extremes and the
evaluation of week-3 subseasonal prediction skill presented in Section
5.

\section{3. Data Sources and
Preprocessing}\label{data-sources-and-preprocessing}

In this study, we use reanalysis and ECMWF hindcast datasets to
construct weekly extremes and event-based metrics for AHT, HR, and MHW
over East Asia. Details of the datasets, including variable names,
period, domain, frequency, resolution, and sources, are listed in Table
5.

{\footnotesize
\setlength{\tabcolsep}{3pt}
\renewcommand{\arraystretch}{1.15}
\def\LTcaptype{table}

\begin{longtable}{@{}
  >{\raggedright\arraybackslash}p{0.13\linewidth}
  >{\raggedright\arraybackslash}p{0.07\linewidth}
  >{\raggedright\arraybackslash}p{0.10\linewidth}
  >{\raggedright\arraybackslash}p{0.20\linewidth}
  >{\raggedright\arraybackslash}p{0.20\linewidth}
  >{\raggedright\arraybackslash}p{0.30\linewidth}@{}}

\caption{Overview of reanalysis and hindcast datasets used in this study.}
\label{tab:data_sources}\\

\toprule
\textbf{Dataset} & \raisebox{-0.5\height}{\shortstack[c]{\textbf{Vari-}\\\textbf{ables}}} & \textbf{Period} &
\textbf{Original domain and T/P resolution} &
\textbf{Processed domain and T/P resolution} &
\textbf{Data access} \\
\midrule
\endfirsthead

\toprule
\textbf{Dataset} & \raisebox{-0.5\height}{\shortstack[c]{\textbf{Vari-}\\\textbf{ables}}} & \textbf{Period} &
\textbf{Original domain and T/P resolution} &
\textbf{Processed domain and T/P resolution} &
\textbf{Data access} \\
\midrule
\endhead

\bottomrule
\endlastfoot

ERA5 &
T2M, TP, SST &
1940--2024 &
\multirow[c]{2}{=}{\raggedright
Global; daily fields on a \(0.25^\circ\) grid} &
\multirow[c]{2}{=}{\raggedright
East Asia (\(21^\circ\)--\(48^\circ\)N, \(114^\circ\)--\(141^\circ\)E);
daily and weekly products on a \(1.5^\circ\) grid} &
\href{https://cds.climate.copernicus.eu/datasets/derived-era5-single-levels-daily-statistics}{Copernicus CDS} 
(\url{https://cds.climate.copernicus.eu/datasets/derived-era5-single-levels-daily-statistics})\\
NOAA OISST &
SST &
1982--2024 &
&
&
\href{https://www.ncei.noaa.gov/products/optimum-interpolation-sst}{NOAA OISST} (\url{https://www.ncei.noaa.gov/products/optimum-interpolation-sst})\\
\addlinespace

ECMWF S2S hindcast V2016 &
\multirow[c]{2}{=}{\raggedright T2M, TP, SST} &
1996--2015 &
Global; daily fields on a \(1.5^\circ\) grid &
\multirow[c]{2}{=}{\raggedright
East Asia (\(21^\circ\)--\(48^\circ\)N, \(114^\circ\)--\(141^\circ\)E);
lead-week-3 products on a \(1.5^\circ\) grid} &
\multirow[c]{2}{=}{\raggedright
\href{https://ecds.ecmwf.int/datasets/s2s-reforecasts?tab=download}{ECMWF ECDS S2S reforecasts}
\url{https://ecds.ecmwf.int/datasets/s2s-reforecasts?tab=download}}
\\

ECMWF S2S hindcast V2024 &
&
2004--2023 &
Global; 6-hourly fields on a \(1.5^\circ\) grid &
&
\\[4pt]

\end{longtable}
}

\subsection{3.1 Reanalysis Data}\label{reanalysis-data}

We used ERA5 daily reanalysis data as the primary reference dataset for
climate extreme events. ERA5 provides a global, long-term record from
1940 to the present, with single-level variables on a regular 0.25° grid
and hourly resolution. From ERA5, we extracted T2M, TP, and SST to
define AHT, HR, and MHW, respectively. MHWs were also defined using NOAA
OISST, which is a daily global SST analysis product that combines
satellite images with in situ measurements from ships and buoys,
providing 0.25° spatial resolution, a continuous record from September
1981 onward, and a gridded SST field.

As the original data differ in spatial and temporal resolution and data
structure, all data are remapped to a common East Asia domain and a
unified space-time grid. The final analysis domain spans 21°--48°N and
114°--141°E, and all variables are cropped to this domain and
subsequently regridded to a regular 1.5° grid to ensure spatial
consistency among variables, using bilinear interpolation for T2M and
SST and an area-conservative method for TP. Daily fields are then
aggregated to weekly means (T2M, SST) or accumulations (TP), and
climatological means and quantile-based thresholds are calculated
separately for the daily and weekly products on the common domain and
grid. AHT and HR are evaluated over all grid points in the analysis
domain, whereas MHW is evaluated only over ocean grid points.

To obtain consistent anomalies and extreme thresholds, we adopt
1991--2020 as the reference climatological period. For this period, we
compute daily and weekly climatological mean fields and subtract the
corresponding day-of-year climatology (for weekly data, the climatology
associated with the week's start date) to calculate daily and weekly
anomalies. For extremes, we estimate the 90th and 95th percentiles from
the same period and use them as thresholds. These procedures are applied
consistently to the ECMWF hindcast data, enabling comparison of extreme
prediction skill between reanalysis datasets and model hindcasts under a
common analysis framework.

\subsection{3.2 ECMWF Hindcast Data}\label{ecmwf-hindcast-data}

The ECMWF Integrated Forecasting System (IFS) S2S hindcast dataset is an
ECMWF reforecast dataset available through the S2S Prediction Project
and is widely used to assess predictability from medium-range to
subseasonal time scales. Here, ``hindcast'' refers to retrospective
forecasts initialized for past dates and used to estimate model
climatology and forecast skill. In this study, T2M, TP, and SST are also
extracted from the ECMWF S2S hindcast over the East Asia domain. We used
both Version 2016 (V2016) and Version 2024 (V2024) of the ECMWF S2S
hindcast, which provide complementary temporal coverage. We applied
analogous spatial and temporal preprocessing to place the hindcasts on
the common analysis grid and temporal framework. Each variable was
cropped to the East Asia domain, bilinearly regridded to a common 1.5°
grid, and aggregated from daily or 6-hourly fields into weekly means
(T2M, SST) and totals (TP).

\section{4. Climate Extremes and Events}\label{climate-extremes-and-events}

\subsection{4.1 Definition of Daily and Weekly Extremes}\label{definition-of-extreme}

To provide a consistent framework across the three extreme phenomena
considered in this study, Anomalously High Temperature (AHT), Heavy
Rainfall (HR), and Marine Heatwaves (MHW), we first introduce a unified
notation. Let \(X_d(t, x, y)\) and \(X_w(t, x, y)\) denote the daily and
weekly values of a climate variable X---representing T2M, TP, or SST---at a grid point (x, y) and time step t. The corresponding seasonally
adjusted extreme thresholds at daily and weekly scales are denoted
\(X^{\text{thr}}_d(\text{doy}, x, y)\) and
\(X^{\text{thr}}_w(\text{doy}, x, y)\), where doy denotes the day of
year associated with the daily value or the start date of the 7-day
window, respectively. These thresholds are computed as the 90th or 95th
percentile of the raw variable over a 30-year climatological baseline
from 1991 to 2020. Daily thresholds are calculated as follows: For each
calendar day, all values falling within a centered sampling window of ±5
days (11 days in total) across all years of the baseline period are
first collected, and the percentile is then computed from these pooled
samples. The resulting day-of-year percentile series is then smoothed
with a ±15-day moving average to yield a seasonally continuous and
stable threshold that accounts for the gradual evolution of the seasonal
cycle. For weekly thresholds, daily fields are first aggregated to
weekly values using a 7-day moving window: moving averages for T2M and
SST and moving sums for TP, to obtain \(X_w(t, x, y)\). To construct
weekly thresholds over the climatological baseline, we collect the 7-day
aggregated values starting on a given day-of-year and at the target date
and four neighboring offsets (-4, -2, 0, +2, +4) across all years of the
baseline period, thereby assembling a locally sampled set of weekly
values that captures short-term variability around the target date. From
these samples, the 90th and 95th percentiles are computed to define
\(X^{\text{thr}}_w(\text{doy}, x, y)\), ensuring that weekly extreme
thresholds are consistent with the daily-scale definition while
remaining seasonally smooth and statistically robust.
The signed departure from the extreme threshold is defined at both
daily and weekly scales as:
\begin{align}
X^e_d(t,x,y)
&= X_d(t, x, y)
 - X^{\text{thr}}_d(\text{doy}, x, y),
\label{eq:daily-extremeness} \\
X^e_w(t,x,y)
&= X_w(t,x,y)
 - X^{\text{thr}}_w(\text{woy}, x, y),
\label{eq:weekly-extremeness}
\end{align}
where positive values indicate conditions exceeding the extreme
threshold.
AHT is identified when the daily or weekly mean 2-meter air temperature
exceeds its seasonally adjusted percentile threshold, i.e., \(T2M^e_d\)
\textgreater{} 0 or \(T2M^e_w\) \textgreater{} 0. HR is defined when the
daily or weekly accumulated precipitation surpasses the 90th or 95th
percentile threshold, i.e., \(TP^e_d\) \textgreater{} 0 or \(TP^e_w\)
\textgreater{} 0. MHW occurs when the daily or weekly mean SST exceeds
its seasonally adjusted threshold, i.e., \(SST^e_d\) \textgreater{} 0 or
\(SST^e_w\)\textgreater{} 0. Extreme conditions can be represented
either as binary labels (0: non-extreme, 1: extreme) or as continuous
severity scores based on the magnitude of the departure \(X^e\), providing
flexibility for both classification and regression-based machine
learning applications.

\subsection{4.2 Extreme-Based Metrics: Regular Indices and Irregular Event Detection}\label{extreme-based-metrics}

To characterize the aggregated behavior of extreme conditions within
regular temporal windows, we define a set of weekly and monthly
extremeness indices. Based on the daily departures defined in Section 4.1, three temporally aggregated
metrics including Extreme Days (ED), Maximum Intensity (MI) and Impact
Factor (IF) are defined for all three extreme types (E $\rotatebox{0}{$\in$}$ \{AHT, HR,
MHW\}) at both weekly and monthly scales (see Table 2). ED is the total
number of days within a given week or month during which the daily
departure \(X^e_d\) exceeds zero, denoted by
\(ED^{\text{E}}_w(\text{t}, x, y)\) and
\(ED^{\text{E}}_m(\text{t}, x, y)\), respectively. MI is the maximum
value of \(X^e_d\) within the window, capturing peak intensity, denoted
by \(MI^{\text{E}}_w(\text{t}, x, y)\) and
\(MI^{\text{E}}_m(\text{t}, x, y)\). A cumulative exceedance metric,
here termed IF, is the the cumulative sum of \(X^e_d\) over all days
within the window, representing the total impact or severity of the
extreme episode, expressed as \(IF^{\text{E}}_w(\text{t}, x, y)\) and
\(IF^{\text{E}}_m(\text{t}, x, y)\).

In addition to the temporally regular indices described above, discrete
extreme ``events'' are identified at the daily level following the framework
of Hobday et al.~(2016). A contiguous period during which
\(X^e_d(t,x,y)\) \textgreater{} 0 is classified as a single event if it
persists for at least n consecutive days, with gaps of up to m days
between exceedances permitted to account for brief interruptions. n and
m denote the minimum event duration and the maximum allowable gap
between threshold exceedances, respectively. We adopt event detection
criteria of D3G5 (minimum 3-day duration, maximum 5-day gap) and D5G2
for AHT and MHW, and D1G3 and D3G3 for HR, reflecting the differing
temporal characteristics of atmospheric versus oceanic extremes. For
each detected event at every grid point, a set of extreme properties are
computed, including onset date, end date, total duration, peak
intensity, and cumulative impact. This is a broadly used procedure to
detect and classify weather extremes (Fischer \& Schär, 2010; Stefanon
et al., 2012; Perkins \& Alexander, 2013; Westby et al., 2013; Nairn \&
Fawcett, 2015; Russo et al., 2015). The event detection algorithm is
adapted from the open-source MHW detection code developed by Oliver et
al.~(available at \url{https://github.com/ecjoliver/marineHeatWaves}),
which has been modified to accommodate the generalized threshold
framework and variable types used in this study.

\section{5. Statistical Analyses of Extreme Events and Weekly Extremes}\label{statistical-analysis-of-extreme-events-and-weekly-extremes}

This section presents the event statistics and weekly extreme features derived using the exemplary codes and data provided by this~repository.
These results facilitate a comprehensive understanding of the historical
occurrence characteristics of AHT, HR, and MHWs in both reanalysis
datasets and ECMWF models.

\subsection{5.1 Daily Timeseries and Extreme Event Detection}\label{daily-timeseries-and-extreme-event-detection}

Figure 2 illustrates how AHT events are isolated from the daily time
series of T2M at 36.0°N, 127.5°E in the central-western region of South
Korea for 2024, based on the corresponding percentile threshold and
detection criteria. The seasonally varying threshold
\(T2M^{\text{thr}}_d(\text{doy}, x, y)\) yields the smooth red curve that
closely tracks the seasonal envelope of temperature. Days on which
\(T2M^e_d(t, x, y) = T2M_d(t, x, y) - T2M^{\text{thr}}_d(\text{doy}, x, y) > 0\)
are flagged as extreme (gray shading). These flagged days are
subsequently passed through the D3G5 event criterion---requiring at
least 3 consecutive exceedance days with gaps of no more than 5 days---to consolidate isolated spikes into physically coherent AHT events (red
shading). As seen in the figure, episodic warm anomalies in spring and
summer tend to remain as isolated extremes, while the sustained warm
period in late summer and autumn qualifies as a prolonged AHT ``event'',
demonstrating the added value of the event-based classification over
simple daily threshold exceedance. Figure 3 demonstrates the timeseries
for TP in the same manner as in Figure 2, but using the event detection
criterion D1G3. Days on which the exceedance
\(\mathrm{TP}^e_d(t, x, y) = \mathrm{TP}_d(t, x, y) - \mathrm{TP}^{\text{thr}}_d(\mathrm{doy}, x, y)\)
is positive are designated as daily extremes (gray shading). These
classified days are further refined using the D1G3 event criterion to
merge closely spaced extreme spikes into HR events (red shading).
Compared with the more stringent D3G5 criterion used for AHTs, the D1G3
criterion captures a larger number of short-lived, intermittent HR
episodes associated with intense precipitation cases, thereby offering a
complementary perspective on extreme precipitation extremes. Both
figures 2 and 3 were generated using the exemplary code
\path{Extreme_Event_Statistics_and_Visualization_1.ipynb} presented
in Section 2B. All the other figures can be found on the repository.

\begin{figure}[htbp]
\centering
\pandocbounded{\includegraphics[keepaspectratio]{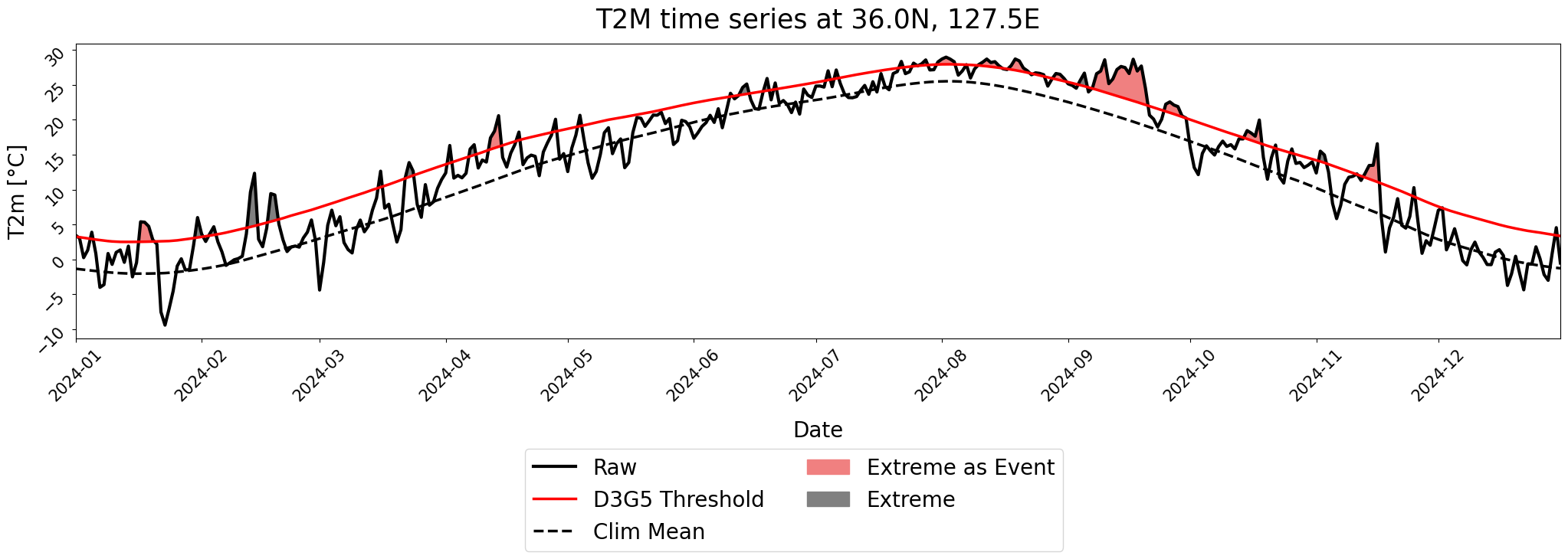}}
\caption{Timeseries of AHT extreme detection at a representative grid
point (36.0$^\circ$N, 127.5$^\circ$E) for 2024. The thick black line
shows the daily T2M time series (\(T2M_d\)), the red line represents
the seasonally varying 90th-percentile threshold
(\(T2M^{\mathrm{thr}}_d\)), and the black dashed line shows the daily
climatological mean. Gray shading indicates individual days for which
\(T2M^e_d > 0\), whereas red shading denotes discrete AHT events
identified using the D3G5 criterion (minimum duration of 3 days and a
maximum allowable gap of 5 days).}
\label{fig:aht_daily_timeseries_2024}
\end{figure}

\begin{figure}
\centering
\pandocbounded{\includegraphics[keepaspectratio]{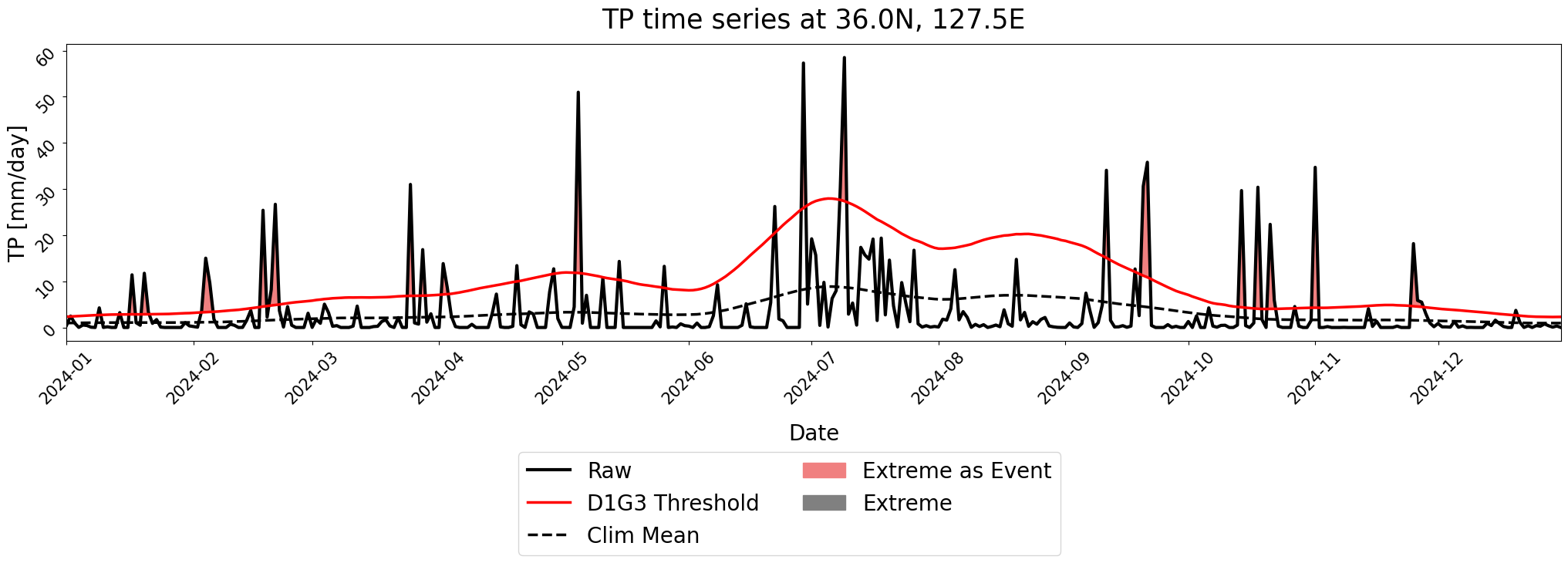}}
\caption{Timeseries of HR extreme detection at a
representative grid point (36.0°N, 127.5°E) for the year 2024.
Everything is the same as in Figure 2, but the event detection criterion
that is D1G3 (minimum 1 consecutive day, maximum 3-day gap).}
\label{fig:hr_daily_timeseries_2024}
\end{figure}

\subsection{5.2 Observed Event
Statistics}\label{observed-event-statistics}

Extreme ``events'' were identified using a 90th percentile threshold at
each grid point within the East Asia domain, enabling a detailed
analysis of the statistical characteristics of occurrence frequencies
and the spatial patterns of events profiles. Figure 4 illustrates the
annual counts of AHT events at 36.0°N, 127.5°E from 1940 to 2024. The
time series delineates a clear upward trajectory, directly reflecting
the global warming signal and the increasing frequency of hot days.
Furthermore, Figure 5 presents a heatmap of monthly HR event counts from
January 1940 to December 2024 at the same location. The (+) and (-)
markers on the right margin of the heatmap indicate the long-term trend
directions for each individual month. With the exception of May, June,
and December, increasing trends were detected across all months. These
visualizations---both the line plot and the heatmap---were generated
using the sample script
(\path{Extreme_Event_Statistics_and_Visualization_1.ipynb}) detailed in
Section 2B. The figures for the other variables can be found on the
repository.

\begin{figure}
\centering
\pandocbounded{\includegraphics[keepaspectratio]{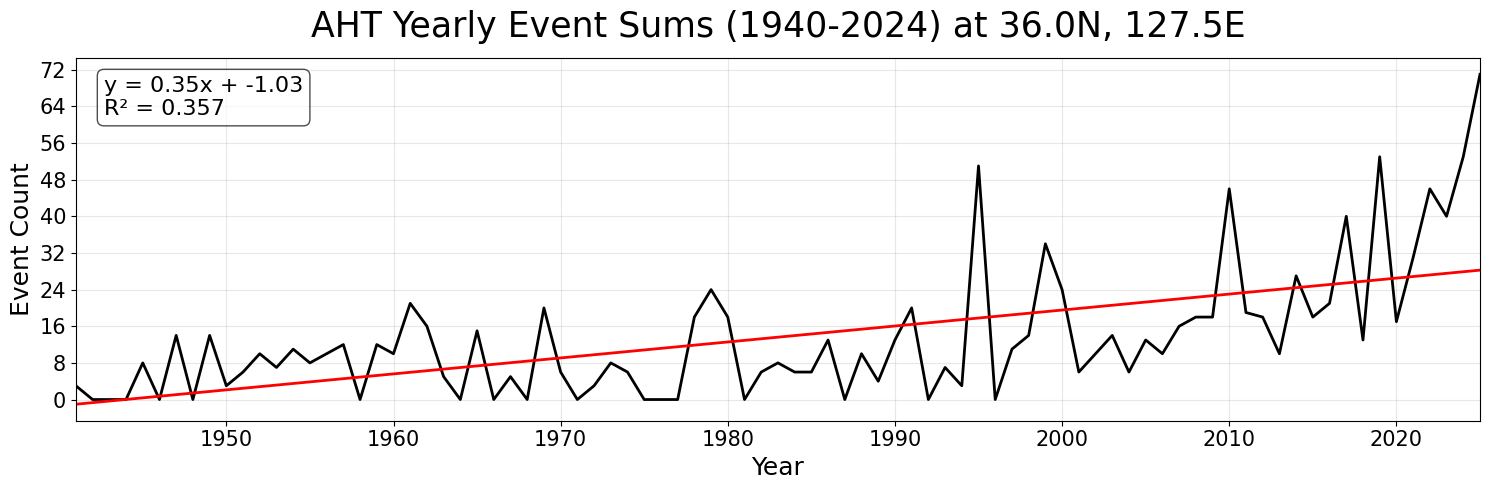}}
\caption{Yearly extreme event sums of anomalously high
temperatures at 36°N, 127.5°E for the period of 1940-2024.}
\label{fig:aht_yearly_event_sums}
\end{figure}

\begin{figure}
\centering
\pandocbounded{\includegraphics[keepaspectratio]{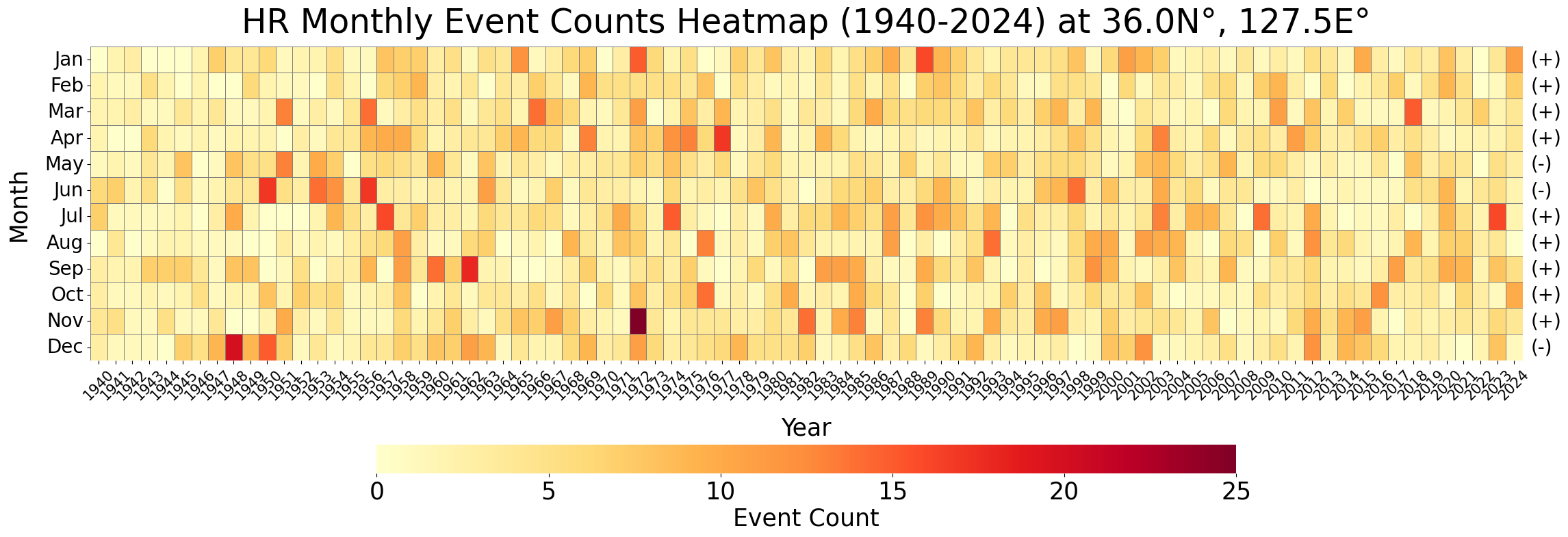}}
\caption{Heatmap of monthly extreme event counts of heavy
rainfall at 36°N, 127.5°E for the period of 1940-2024.}
\label{fig:hr_montly_event_counts_heatmap}
\end{figure}

Figure 6 presents spatial maps of long-term statistics over the East
Asian Marginal Seas---including the climatological mean and decadal
trend of event frequency, the total number of days with anomalously high
SST associated with these events, and average event
intensity---generated using the sample script
(\path{Extreme_Event_Statistics_and_Visualization_2.ipynb}) described in
Section 2B. The applied threshold for extreme criteria is the 90th
percentile, and events are detected using a minimum duration of 5 days
and a maximum gap of 2 days. The upper panels show the average annual
event frequency, total event days, and mean intensity over the
1982--2024 period. The results indicate that MHWs are more pronounced in
marginal seas such as the East Sea, the Yellow Sea, and the East China
Sea than in the open Pacific east of the Japanese archipelago, with MHW
intensity notably stronger in the East Sea and along the Polar Front
region. In the lower panels, a decadal trend analysis based on the
annual statistics at each grid point reveals a robust increasing trend
in both event frequency and total event days. Event mean intensity,
however, does not follow this uniform increasing pattern; instead, it
exhibits a spatially decoupled structure, strengthening in the marginal
seas adjacent to the continent while weakening toward the open Pacific.

\begin{figure}
\centering
\pandocbounded{\includegraphics[keepaspectratio]{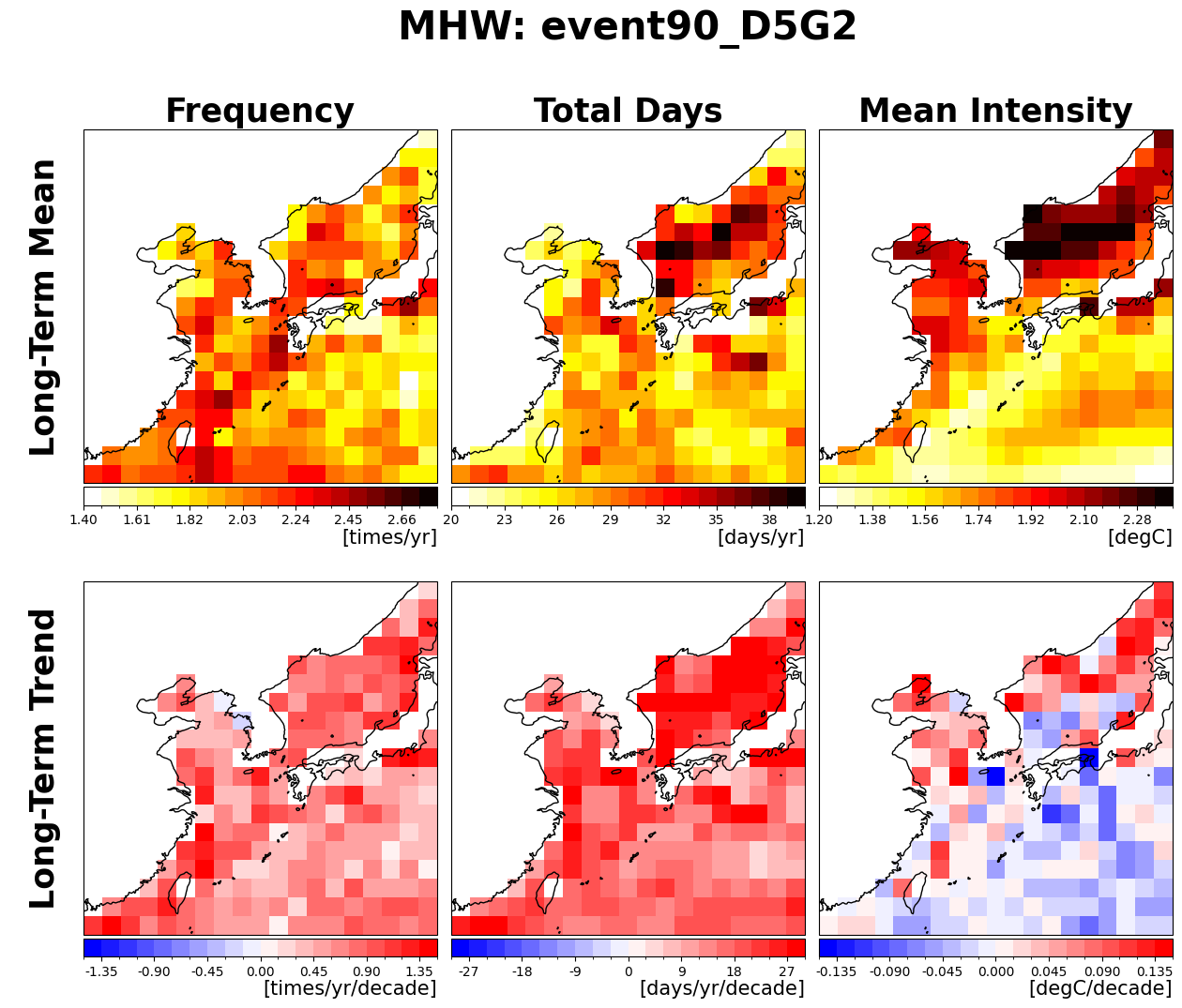}}
\caption{Spatial distributions of long-term mean (top) and
trend (bottom) of MHW event statistics over East Asia during
1940--2024: frequency (left), total days (center), and mean intensity
(right).}
\label{fig:mhw_event_spatial}
\end{figure}

\subsection{5.3 Observed Occurrence and Characteristics of Weekly
Extremes}\label{observed-occurrence-and-characteristics-of-weekly-extremes}

In contrast to the extreme events and extremeness indices discussed earlier in Sections 4.2 and 5.2, both derived from daily data [Eq. (1)], this section examines the extremeness of the weekly representative value [Eq. (2)] and characterizes its statistical properties. 
Figure 7 presents the annual statistics derived from the heavy rainfall
weeks, including the occurrence frequency, mean intensity, and the
annual maximum intensity. While the frequency is notably higher over the
continent than over the ocean, the mean intensity exhibits a distinctly
strong pattern along the Japanese archipelago, tracing the path of the
Kuroshio Current. The spatial distribution of the annual maximum
intensity largely resembles the mean intensity pattern; however, it
tends to appear somewhat noisier due to the large variance in extreme
intensity values. Similarly, Figure 8 illustrates the annual statistics
for the abnormalously high temperature weeks extracted using the same
approach. A distinctly high frequency of AHT occurrences is observed
over South China, the southwestern Korean Peninsula, and the southern
tip of Honshu, Japan. Both the mean and annual maximum intensities
exhibit a clear meridional gradation---higher in the south and lower in
the north. This reflects the inherent climatological pattern where
absolute temperatures are higher at lower latitudes and gradually
decrease toward higher latitudes. Nevertheless, the annual maximum
intensity reveals localized peak values over parts of South China and
the Korean Peninsula, capturing the strong effects of continental land
surface heating. Both figures were generated using the exemplary code,
\path{Weekly_Extreme_Statistics_and_Visualization.ipynb},~presented in
Section 2B. All the other figures can be found on the repository.

\begin{figure}
\centering
\pandocbounded{\includegraphics[keepaspectratio]{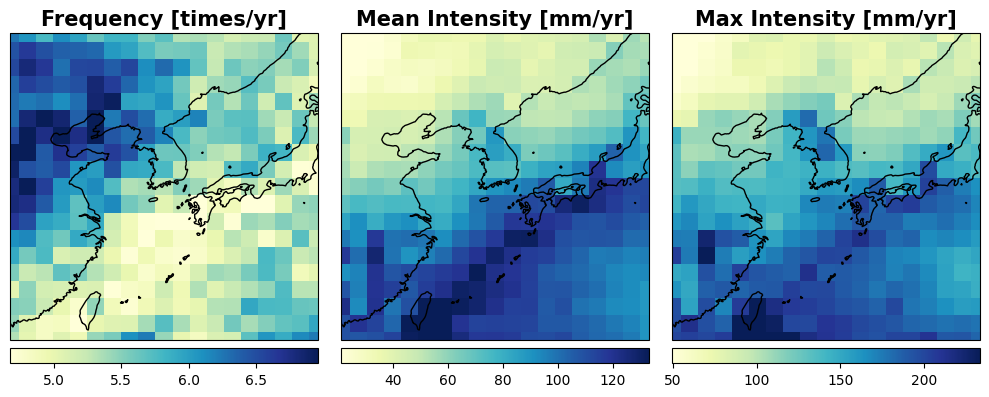}}
\caption{Spatial distributions of long-term weekly heavy
rainfall (HR) statistics over East Asia during 1940--2024: mean annual
frequency (left), mean annual extreme week intensity (middle), and mean
annual maximum intensity (right). Weekly HR was defined as 7-day
accumulated precipitation exceeding a day-of-year (doy)-dependent
90th-percentile threshold based on the 1991--2020 reference period. The
weekly series was sampled on Mondays to represent successive weekly
windows.}
\label{fig:hr_weekly_spatial}
\end{figure}

\begin{figure}
\centering
\pandocbounded{\includegraphics[keepaspectratio]{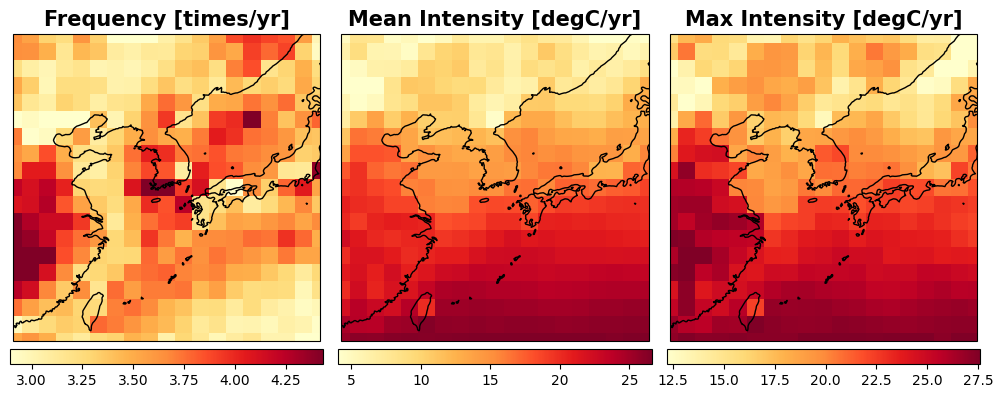}}
\caption{Spatial distributions of long-term weekly anomalously
high temperature (AHT) statistics over East Asia during 1940--2024: mean
annual frequency (left), mean annual extreme week intensity (middle),
and mean annual maximum intensity (right). Weekly AHT was defined as
7-day mean 2-m air temperature exceeding a day-of-year (doy)-dependent
90th-percentile threshold based on the 1991--2020 reference period. The
weekly series was sampled on Mondays to represent successive weekly
windows.}
\label{fig:aht_weekly_spatial}
\end{figure}

\subsection{5.4 ECMWF Hindcast Reproducibility of Weekly Extremes}\label{ecmwf-hindcast-weekly-extremes-reproducibility}

In this study, we compare the occurrence characteristics of each extreme
component derived from ECMWF S2S hindcasts (V2016 and V2024) and ERA5
reanalysis, in order to evaluate the model's prediction skill and bias
compared to the observation-based reference. First, for both ECMWF
hindcasts and ERA5, we compute the frequency with which 7-day
running-mean values at lead week 3 exceed the version-specific
90th-percentile climatological threshold, aggregate these exceedance
counts by year, and then average over the hindcast periods to obtain
maps of annual frequency of extreme weeks over East Asia. This analysis
allows us to assess how consistently the ECMWF hindcasts reproduce the
observed frequency, intensity, and spatiotemporal structure as
diagnosed from ERA5, and further to evaluate the changes in prediction
skill and bias structure between V2016 and V2024.

Figure 9 compares the spatial distribution of the annual mean frequency
of AHT occurrence (weeks per year), as represented by the ERA5
reanalysis and the ECMWF hindcast model, for two hindcast periods
(1996--2015 and 2004--2023). In the first period (first and second
columns), both figures exhibit a relatively uniform distribution of
heatwave frequency of about 4--5 weeks per year, with only small
inter-grid differences and broadly consistent spatial patterns between
ERA5 and the ECMWF hindcast. In contrast, in the second period (third
and fourth columns), the frequency increases to more than 6--7 weeks per
year, indicating an overall intensification of AHT occurrence. However,
it is noteworthy that the spatial structure differs markedly between the
reanalysis and the model: ERA5 shows enhanced frequencies over eastern
China and the Yellow Sea, whereas the ECMWF hindcast indicates stronger
occurrence over higher-latitude continental regions, including Manchuria
and eastern Russia. These discrepancies can be attributed to inherent
limitations of dynamical models, such as incomplete representation of
Earth system interactions, imperfect physical parameterizations, and
reduced skill in simulating extremes. In addition, the relatively narrow
sample distribution/spread of high-latitude continental temperatures in the
dynamical model, compared to reanalysis, likely leads to increased
sensitivity of the estimated quantiles, thereby amplifying differences
in the diagnosed heatwave frequencies.

Figure 10 shows the temporal evolution of weekly temperature and
associated extremes in 2023 at 36.0°N, 127.5°E in the central-western
region of South Korea, from the ECMWF model and ERA5 observations. The weeks during which ERA5 temperatures (red solid line) exceed the corresponding 90th-percentile threshold (Clim90; red dashed line) are highlighted in red shading, while weeks during which the ECMWF hindcast (black solid line) exceeds its threshold (gray dashed line) are indicated by gray shading. Overall, the seasonal cycle of temperature in
2023 is broadly consistent between the ECMWF model and ERA5, but the
model exhibits a cold bias, particularly during the first half of the
year, and thus fails to reproduce the anomalously warm annual mean
temperature observed in 2023. Moreover, substantial discrepancies emerge
in the simulation of heat extremes. In early March, late October, and
early December, when ERA5 temperatures exceed the Clim90 threshold and
distinct hot extremes are observed (red shaded periods), the ECMWF
hindcast either remains below the threshold or shows a delayed onset of
high temperatures. These results suggest that medium-range to
subseasonal (weekly-scale) prediction models fall short of capturing abrupt temperature fluctuations and heatwave events.

\begin{figure}[htbp]
\centering
\pandocbounded{\includegraphics[keepaspectratio]{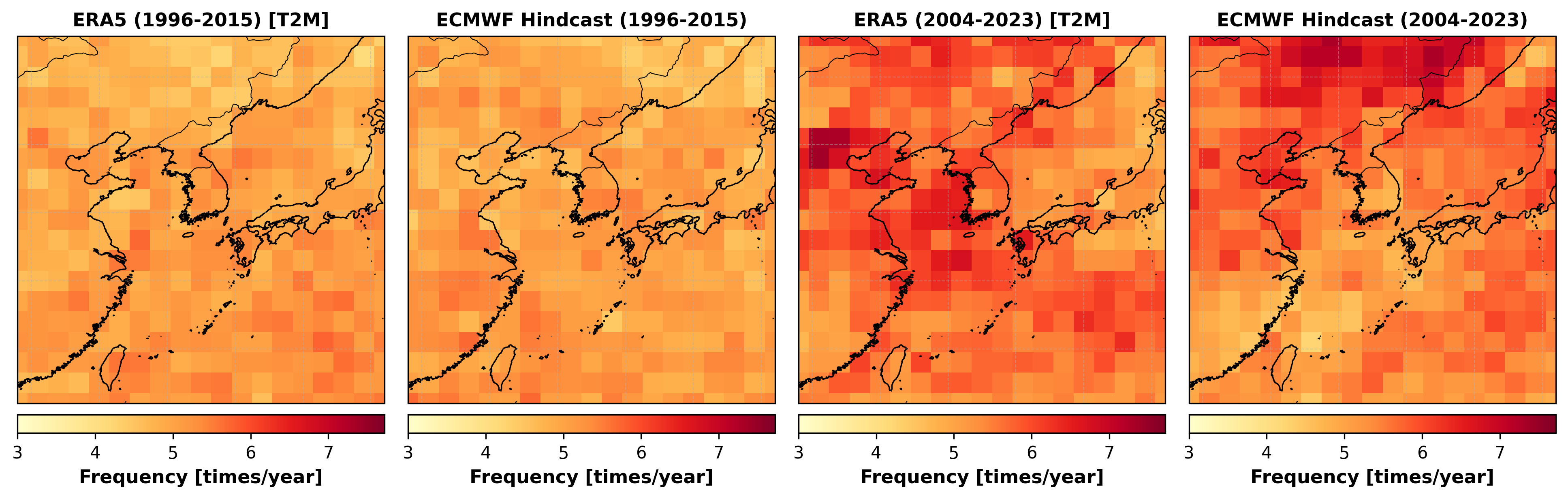}}
\caption{Spatial distribution of the mean annual frequency of
anomalously high temperature (AHT) weeks derived from ERA5 reanalysis
and ECMWF hindcast over two 20-year periods: 1996--2015 (left panels)
and 2004--2023 (right panels). For the ECMWF hindcast, only lead week 3
forecasts are used, from which weekly mean time series are constructed
prior to AHT detection.}
\label{fig:ecmwf_era5_weekly_map}
\end{figure}

\begin{figure}[htbp]
\centering
\pandocbounded{\includegraphics[keepaspectratio]{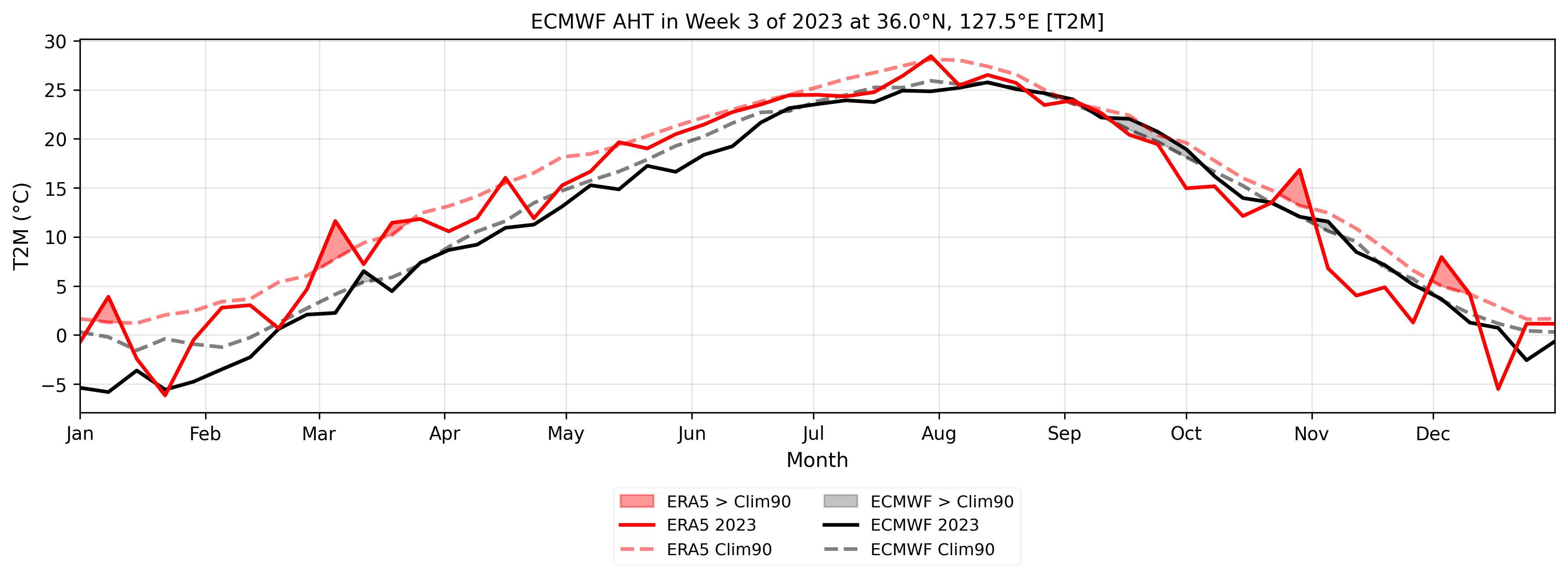}}
\caption{Time series of weekly mean temperature and
anomalously high temperature (AHT) weeks at 36.0°N, 127.5°E for the year
2023. Solid red and black lines represent ERA5 reanalysis and ECMWF
hindcast (V2024), respectively. Dashed red and gray lines indicate the
climatological mean and 90th percentile threshold for ERA5 (2004--2023)
and ECMWF hindcast (2004--2023), respectively. Red and gray shading
denotes weeks in which the weekly mean temperature exceeded the
corresponding 90th percentile threshold, identifying AHT weeks in ERA5
and ECMWF, respectively. For the ECMWF hindcast, lead week 3 forecasts
are extracted and aggregated into weekly mean time series prior to AHT
detection.}
\label{fig:ecmwf_era5_weekly_timeseries}
\end{figure}

\FloatBarrier
\section{6. Conclusion}\label{conclusion}

The EastAsiaClimateExtremes repository provides the open dataset and
analysis workflows described in this paper, which is hosted on GitHub.
The repository is structured to help users easily access data, reproduce
key results, and adapt scripts for custom applications. This dataset
offers a structured inventory of grid-based weekly extreme labels and
event-based metrics over East Asia. It covers three main climate
extremes: anomalously high temperature (AHT), heavy rainfall (HR), and
marine heatwaves (MHW). Reanalysis-based extreme definitions are
explicitly documented. Crucially, these are co-registered with ECMWF S2S
hindcast outputs on a common grid and temporal framework. The
observation-derived labels are formatted to be directly comparable to
dynamical model forecasts. As a result, the dataset can serve as a
standardized reference for East Asian climate extreme research and
AI-based subseasonal-to-seasonal prediction experiments. Researchers can
use this gridded extremes dataset to quantitatively characterize the
spatiotemporal occurrence of regional extremes. Furthermore, the direct
alignment with ECMWF hindcasts could enable a systematic diagnosis of
the model's strengths and limitations in reproducing extreme signals.
Beyond these immediate uses, the dataset has the potential to function
as a common foundation for broader studies. These might include
subseasonal forecast benchmarking, extreme event attribution, and
compound risk analysis. 

Several limitations should be acknowledged. The
reanalysis and analysis products underpinning the dataset---ERA5 and
OISST---may carry biases inherited from observational network
constraints, data assimilation procedures, and model dependence. These
uncertainties can propagate into the anomaly fields and percentile
thresholds used to define extremes. Accordingly, the resulting extremes
should not be interpreted as directly equivalent to those derived
strictly from direct physical measurements, such as station-based or
in-situ ocean observations. The ECMWF S2S hindcast is similarly subject
to systematic errors, including underestimation of precipitation
extremes and SST biases, which warrant caution when interpreting HR and
MHW prediction skill. The extreme definitions themselves carry a degree
of sensitivity to methodological choices, such as the selection of the
90th versus 95th percentile threshold, sampling and smoothing windows
for the thresholds, or the minimum persistence criterion, and results
may vary accordingly. While the current labels and indices effectively
summarize extreme characteristics within regular temporal windows at the
grid-point level, they present limitations for AI applications that
require event-level profiles---including onset, duration, spatial
propagation, and morphology of individual episodes. Although a simple
extremeness index is provided as a temporally regular proxy of daily
extreme behavior, this falls short of a proper object- or event-level
representation, which remains an open challenge for translating extreme
event characteristics into machine-learning-compatible label structures.

Looking ahead, the label framework can be extended beyond AHT, HR, and
MHW to encompass a broader range of hazard types including drought, cold
surges, strong winds, wave extremes, and compound events. Given the
ERA5, OISST, and ECMWF S2S inputs already in place, additional variables
such as soil moisture, surface runoff, wind speed, and significant wave
height can be incorporated with relatively straightforward
post-processing to derive weekly labels for these hazard types,
ultimately enabling a unified multi-hazard label inventory for East
Asia. The current grid-point-based weekly labels can also be extended
toward finer temporal resolutions and object-based labeling schemes, in
which spatially contiguous extreme regions are treated as discrete
objects with defined boundaries, movement, and lifecycle properties---opening the door to an object-oriented dataset that captures the spatial
structure and propagation of heat domes, rainfall bands, and marine
heatwave patches. The existing alignment between ECMWF S2S hindcasts and
observation-based labels provides a natural scaffold for incorporating
additional model hindcasts from and CMIP6 or decadal prediction outputs,
thereby expanding the dataset into a multi-source ensemble benchmark
that supports multi-model bias correction, model weighting, and AI-based
ensemble prediction experiments. Building upon the preprocessed inputs
currently provided, physics-aware, multi-task, and multi-hazard AI
models can be developed within this framework to simultaneously forecast
and diagnose multiple extreme phenomena and associated atmospheric and
oceanic indices across the East Asian domain. Finally, through regular
updates to data and code, integration of new variables and label
definitions, and incorporation of community feedback, the dataset is
expected to evolve into a sustained public infrastructure for East Asian
climate extreme research and AI-based prediction science. As a next step toward this longer-term development, we aim to incorporate real-time forecast data, enabling the same label framework to support not only retrospective hindcast evaluation but also operational S2S forecast applications.

\section{Acknowledgments}\label{acknowledgments}

This research was funded by the Korea Meteorological Administration
Research and Development Program ``APEC Climate Center for Climate
Information Services'' under Grant (KMA2013-03410). We acknowledge the
Copernicus Climate Change Service for providing the ERA5 reanalysis
data, ECMWF for making the Subseasonal-to-Seasonal (S2S) hindcast data
available through the S2S Prediction Project database, and the NOAA
National Centers for Environmental Information for providing the NOAA
Optimum Interpolation Sea Surface Temperature (OISST) data. This work was also supported by Korea Institute of Science and Technology Information/Korea Research Environment Open NETwork (KISTI/KREONET).

\section{Code and Data
Availability}\label{code-and-data-availability}

The source code and analysis workflows used to reproduce the figures and
statistics presented in this study are publicly available at
\url{https://github.com/yyalexlee/EastAsiaClimateExtremes}. The
EastAsiaClimateExtremes data products are publicly available through
Zenodo at \url{https://doi.org/10.5281/zenodo.22139821}. Instructions
for downloading the data, descriptions of the directory structure, and
file-level metadata are provided in the GitHub repository.

\end{document}